\documentclass[pre,showpacs,twocolumn,floatfix,bibnotes]{revtex4}
\usepackage{here}
\usepackage[dvips]{graphicx}

\newcommand\pictc[5]{\begin{figure}
                       \centerline{
                       \includegraphics[width=#1\columnwidth]{#3}}
                   \protect\caption{\protect\label{fig:#4} #5}
                    \end{figure}            }
\newcommand\pict[4][1.]{\pictc{#1}{!tb}{#2}{#3}{#4}}
\newcommand\rpict[1]{\ref{fig:#1}}
\newcommand\leqt[1]{\protect\label{eq:#1}}

\newcounter{Fig}

\begin{document}

\begin{sloppy}

\title{Giant Goos-H\"anchen effect at the reflection from left-handed metamaterials}

\author{Ilya V. Shadrivov$^1$, Alexander A. Zharov$^{1,2}$, and Yuri S. Kivshar$^1$}

\affiliation{$^1$Nonlinear Physics Group, Research School of
Physical Sciences and Engineering, Australian National
University, Canberra ACT 0200, Australia \\
$^2$ Institute for Physics of Microstructures, Russian Academy of
Sciences, Nizhny Novgorod 603950, Russia}

\begin{abstract}
We study the beam reflection from a layered structure with a left-handed
metamaterial. We predict a giant lateral (Goos-H\"anchen) shift and splitting of the beam due to the resonant excitation of surface polaritons with a vortex-like energy flow between the right- and left-handed materials.
\end{abstract}


\maketitle

As is known, a totally reflected beam experiences a lateral
displacement from its position predicted by the geometric optics
because each of its plane-wave components undergoes a different
phase change~\cite{Goos:1947-333:AP}. For the beam reflected from
an interface, the lateral beam shift (the so-called
Goos-H\"{a}nchen effect) is usually much less than the beam width. However, larger beam shifts may occur in the layered systems
supporting {\em surface waves}, which are able to transfer energy
along the interface. Such surface waves are not excited at a
single interface because the phase-matching condition between the
incident and surface waves are not satisfied. However, a surface polariton can be excited in a layered structure, when the beam is incident at the angle larger than the angle of the total internal reflection. Excitations of polaritons take place in two well known geometries of the attenuated beam reflection (used e.g. for the measurements of the dielectric permittivity of solids): (i)
glass prism-air-silver structure (the so-called Otto
configuration) and (ii) prism-silver film-air structure (the
so-called Kretchmann configuration)~(see, e.g.,
Ref.~\cite{Chuang:1986-593:JOSA} and references therein).

The extensive study of novel microstructured materials with
negative refraction recently fabricated
experimentally~\cite{shelby}, the so-called {\em left-handed
metamaterials}, demonstrates that an interface between right- and
left-handed material supports {\em surface polaritons} of
TE and TM types~\cite{Ruppin:2000-61:PLA,Shadrivov:unpub}, which can enhance dramatically the Goos-H\"anchen effect because
surface waves can transfer the energy along the interface. In this Letter, we predict and analize the giant Goos-H\"{a}nchen effect in the structure (i) (the Otto configuration), where the third medium is substituted by a left-handed metamaterial. 

We consider a two-dimensional geometry shown in Fig.~\rpict{geom_angle}(a), where the beam, incident from an optically dense medium ($\epsilon_1
\mu_1> \epsilon_2 \mu_2$) at the angle larger than the angle of
the total internal reflection,  is reflected from the interface. We assume that the third medium is separated by a gap (which we call the second medium), and it is made of a left-handed metamaterial which possesses both negative real parts of dielectric permittivity $\epsilon_3$ and magnetic permeability $\mu_3$. The interface between the second and third media supports surface waves which can be excited resonantly when the tangential
component of the wave vector of the incident beam coincides with
the propagation constant of the surface polariton. In such a case,
the surface wave can transfer energy along the interface
leading to a great enhancement of the lateral beam shift.

The Goos-H\"{a}nchen shift $\Delta$ can be calculated analytically as $\Delta = -d \Phi/dk_x$~\cite{Brekhovskikh:Waves}, when the phase $\Phi$ of the reflection coefficient is a linear function of the wave vector component $k_x$ along the interface, across the spectral width of the beam. The (standard) lateral beam shift has been calculated for the reflection from layered structures with left-handed materials, for the cases of a single interface~\cite{Berman:2002-067603:PRE} and for a periodic structure of alternating right- and left-handed layers~\cite{Shadrivov:2003-:APL}. 

In our geometry [see Fig.~\rpict{geom_angle}(a)], the reflection coefficient $R= R(k_x)$ for the TE-polarized plane monochromatic [$\sim\exp{(i\omega t)}$] wave can be found as
\begin{equation} \leqt{reflection3}
R = \frac{( \alpha_1 + 1 )(\alpha_2 + 1 )
            -(\alpha_1 - 1 )( \alpha_2 - 1 )e^{2ik_2d}}{ 
			(\alpha_1 - 1 )(\alpha_2 + 1 )
            -(\alpha_1 + 1)(\alpha_2 - 1 )e^{2ik_2d}},
\end{equation}
where $\alpha_{1,2} = k_{z1,3} \mu_2 /k_{z2} \mu_{1,3}$, $k_{zi} =
(\omega^2 \epsilon_i \mu_i /c^2 - k_x^2)^{1/2}$, where $i =1,2,3$,
$d$ is the gap thickness, and $c$ is the speed of light in
vacuum. Here, we consider the TE polarized waves, but the results
are qualitatively similar for the TM polarized waves.

We consider an incident beam of the Gaussian shape, $E_i(x) =
\exp{(-x^2/4a^2 -i k_{x0} x)} $, where $a$ is the width of the
beam, $k_{x0}$ determines the angle of incidence $\phi$, $k_{x0} =
k_1 \sin{\phi}$, and $k_1$ is the propagation constant in the fist
medium, $k_1 = \omega \sqrt{(\epsilon_1 \mu_1)}/c$. The reflected
field is found in a standard way
\begin{equation}
\leqt{reflection_field}
E_r(x) = \frac{1}{2\pi}\int_{-\infty}^{\infty} R(k_x) \bar{E}_i(k_x) \, dk_x,
\end{equation}
where $\bar{E}_i$ is the Fourier spectrum of the incident beam.

\pict{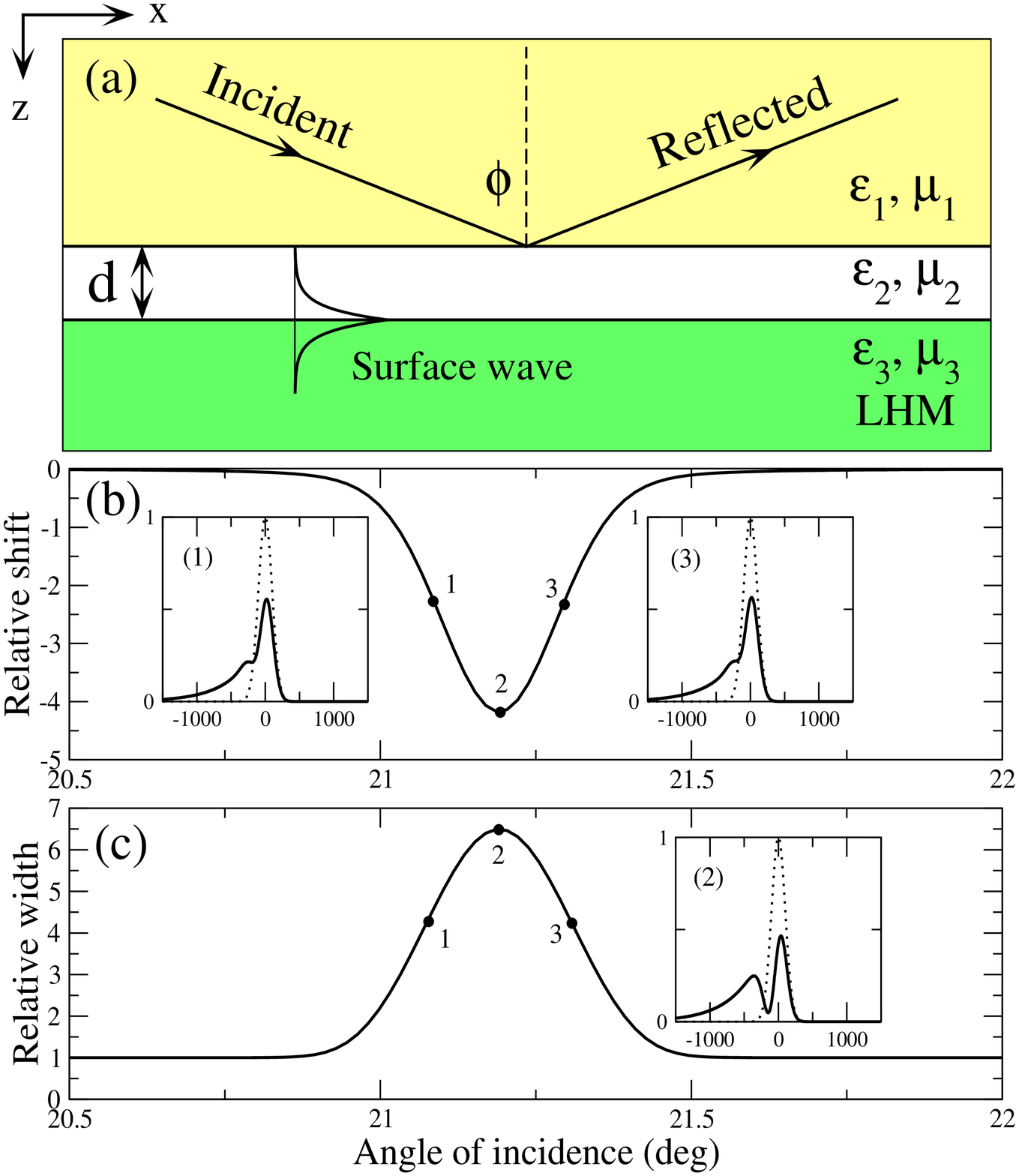}{geom_angle}{ (a) Geometry of the problem. (b,c)
Relative beam shift and width vs. incidence angle.
Insets show the profiles of the reflected (solid) and incident
(dotted) beams.}
\pict{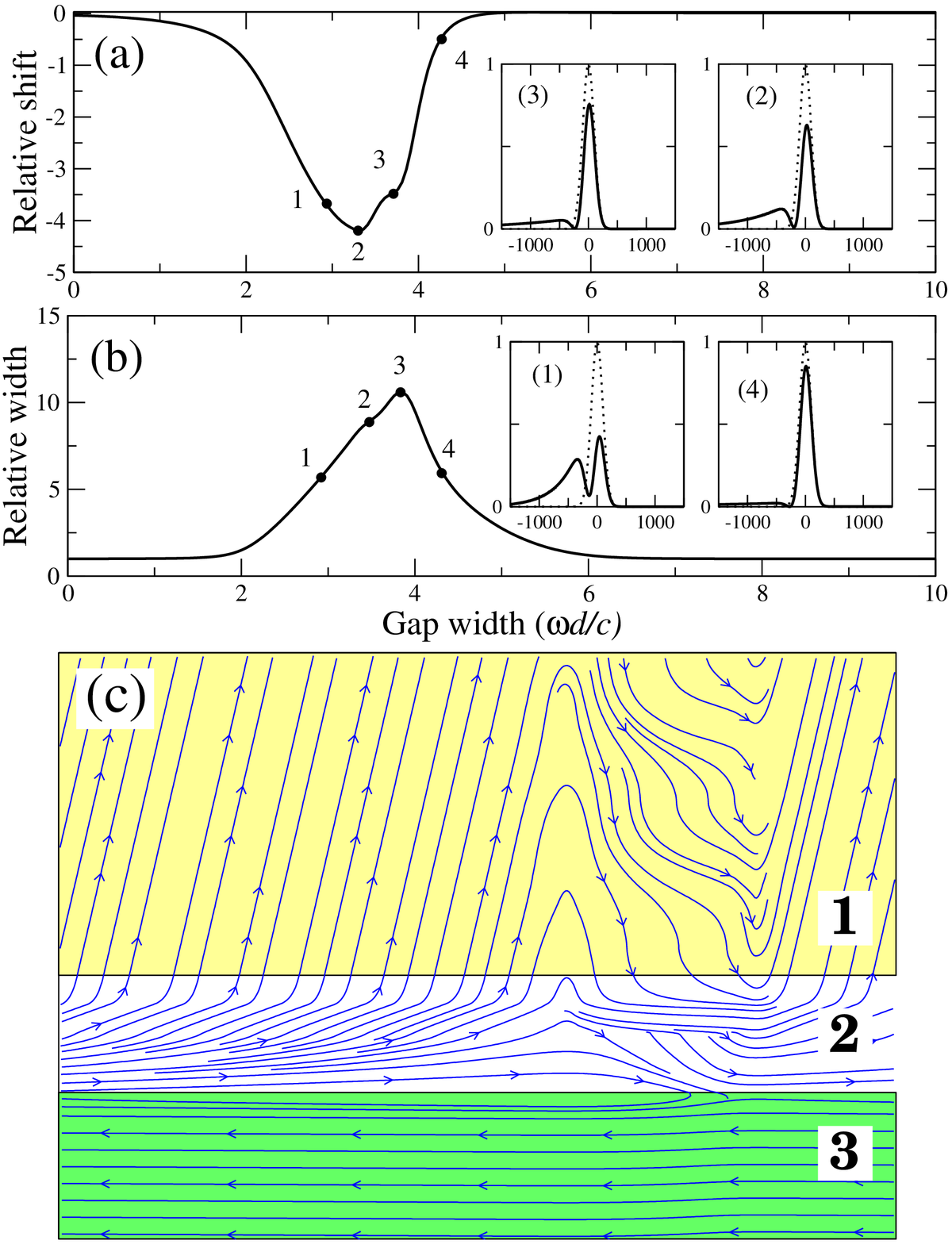}{gap_flow}{ (a,b) Relative shift and width of the reflected beam vs. normalized gap $d \omega/c$ at $a \omega/c = 100$. (c) Energy flow. In all plots the angle of incidence corresponds to the point (2) in Figs.~\rpict{geom_angle}(b,c). Insets show the profiles of the reflected (solid) and incident (dotted) beams.}
\pict{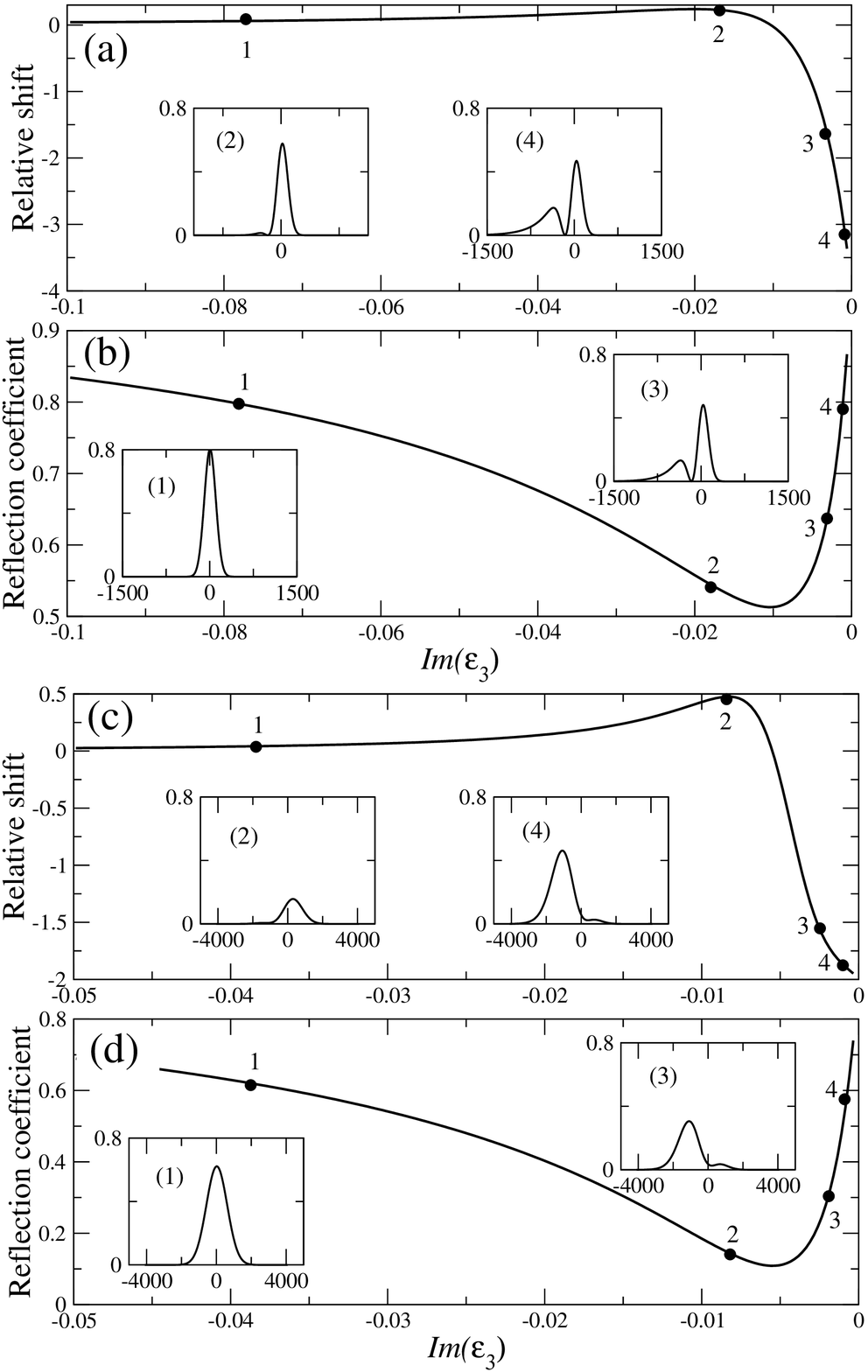}{losses}{ (a,c) Relative beam shift and (b,d)
reflection coefficient vs. the imaginary part of the dielectric
permittivity, for narrow $d\omega/c=100$ (a,b) and wide
$d\omega/c=600$ (c,d) beams, for $d \omega/c = 3$. Insets show the
profiles of the reflected beam.}

Larger lateral beam shifts are expected at the angles of incidence such that the beam spectrum contains the components with the same $k_x$ as the propagation constant of the surface waves. As has been shown recently, both {\em forward and backward} surface polaritons can exist at the left-handed interface~\cite{Shadrivov:unpub}, depending on the parameters $X =|\epsilon_3|/\epsilon_2$ and $Y = |\mu_3|/\mu_2$. Excitation of the forward surface waves results in the energy transfer in the direction of incidence. A negative shift of the beam is obtained when the backward surface waves are excited, this corresponds to the parameters $XY>1$ and $Y<1$.

We define the relative beam shift $\Delta$ as the normalized
{\em first momentum} of the electric field of the reflected beam,
$\Delta = \Delta_1$, where
\begin{equation} \leqt{shift}
    \Delta_n = a^{-n} \int_{-\infty}^{\infty} x^n |E_r(x)|^2dx
    \left( \int_{-\infty}^{\infty} |E_r(x)|^2 dx \right )^{-1}.
\end{equation}
The values $\Delta \ll 1$ correspond to the beam shift much smaller then the beam width, while $\Delta \ge 1$ corresponds to a giant Goos-H\"{a}nchen shift. The second momentum $\Delta_2$ characterizes a change of the beam width during the reflection, $W= \Delta_2^{1/2}$.

We chose the following parameters of the media: $\epsilon_1 =
12.8$, $\mu_1 = \epsilon_2 = \mu_2 = 1$, $\epsilon_3 = -3$, $\mu_3
= - 0.5$. The propagation constant of surface polaritons can be
found from the relation $h^2 = (\omega/c)^2\epsilon_2 \mu_2
Y(Y-X)/(Y^2-1)$, and for these parameters the surface waves at the interface are backward propagating.  Figures \rpict{geom_angle}(b,c) show the dependence of $\Delta$ and $W$ on the angle of incidence at $\omega a /c =100$ and $\omega d /c =3$, respectively. We observe a distinctive
resonant dependence of the beam shift, and the maximum
corresponds to the phase matching condition $k_{x0} = h$.

In the inset (2) of Fig.~\rpict{geom_angle}(c), two distinctive peaks in the reflected beam are observed. The first peak corresponds to a mirror reflection, while the second peak is shifted from the incident beam and it appears due to the excitation of surface waves. At the resonance, the lateral beam shift becomes larger than the beam width. The double-peak structure appears only for narrow beams, when the beam spectrum is wider than the spectrum width of the surface mode, the later can be found as a width of the resonance shown in Fig.~\rpict{geom_angle}(b). The components of the beam spectrum outside this region are reflected in a mirror-like fashion, while the spectrum components near the resonance transform into a surface wave, being responcible for the second peak in the shifted reflected beam. For wider beams, such that their spectrum is completely falls into the surface mode excitation line, only the shifted peak appears. With an increase of the beam width, though, the relative beam shift decreases due to the fact that the absolute shift of the beam grows slower than the beam width.

Figures~\rpict{gap_flow}(a-c) show the beam shift and width vs.
the normalized gap thickness and the structure of the
energy flow. The resonances observed in Figs.~\rpict{gap_flow}(a,b) have simple physical explanation. Indeed, when the gap is absent ($d=0$) or small, no surface wave is excited, and the beam shift is negligible. Increasing the gap
width, one increases the quality factor of the surface mode, thus
increasing its propagation distance and the reflected beam shift.
Similarly, for large $d$ the surface wave is not excited, and the
reflected beam shift becomes small again.

To understand deeper the physical mechanism for the giant
Goos-H\"{a}nchen shift, we calculate the energy flow for the beam
reflection and compare it with the results for the right-hand
media~\cite{Lai:2000-7330:PRE}. Figure~\rpict{gap_flow}(c) shows
the structure of the energy flow for the case of a negative
lateral shift. The strongly curved flow lines in the upper medium
correspond to the interference of the incident and mirror-like
reflected beams. The finite-extension surface wave excited in a
slab along the interface has a distinctive vortex-like
structure~\cite{Shadrivov:PRE}. This surface wave transfers the
energy in the negative direction, and then the energy is reflected
from the interface as a shifted beam.

To make our predictions more realistic, we study the effect of
losses always present in left-handed materials. We introduce
losses by adding the imaginary parts to the dielectric
permittivity $\epsilon_3$ and magnetic permeability $\mu_3$. In
particular, we take ${\cal I}m(\mu_3)= - 2\cdot 10^{-5}$ and vary the
imaginary part of $\epsilon_3$. We notice that the losses in the
left-handed medium affect mostly the surface waves and, therefore,
the major effect produced by the losses is observed for the
strongly shifted beam component.

We distinguish two limiting cases. When the beam is narrow, i.e. its
spectral width is large, only a part of the beam energy is
transferred to surface waves, while the other part is
reflected. This case is shown is Figs.~\rpict{losses}(a,b). In
this case the increase of losses, i.e. the increase of the
absolute value of ${\cal I}m(\epsilon_3)$, results in the suppression
of the second peak in the reflected beam, which is due to the
surface wave excitation. The growth of the amplitude of the
mirror-like reflected beam can be explained by a detuning from the
resonance between the incident wave and the surface polariton.

For a wide beam with a narrow spectrum, almost all energy of the
beam is transferred into a surface wave and, therefore, the
lateral shift becomes strongly affected by the losses in the
left-handed medium, as shown in Fig.~\rpict{losses}(c).
Figure~\rpict{losses}(d) show the dependence of the reflection
coefficient on the imaginary part of the dielectric permittivity
in the third medium.

In conclusion, we have described the Goos-H\"anchen shift and
splitting of a beam totally reflected from a layered structure
with a left-handed metamaterial. The giant lateral shift of the
beam is explained by the resonant excitation of surface
vortex-like polaritons at the interface between the right- and
left-handed materials. The beam shift can be both positive and
negative, depending on the type of the surface waves excited at the
interface. We believe that the giant enhancement of the lateral shift of the
reflected beam can be employed for direct measurements of the
parameters of left-handed materials.

\end{sloppy}
\end{document}